\documentclass[9pt,conference]{IEEEtran}
\usepackage{dcase2025}

\usepackage{bm} 

\usepackage{inconsolata}

\usepackage{xspace}
\newcommand{\method}{\texttt{DRASDIC}\xspace}
\newcommand{\benchmark}{\texttt{FASD13}\xspace}

\usepackage{microtype}
\usepackage{graphicx}
\usepackage{subfigure}
\usepackage{booktabs} 

\usepackage{makecell}

\usepackage{acronym}
\newacro{FSBSED}{few-shot bioacoustic sound event detection}
\newacro{ML}{machine learning}
\newacro{SNR}{signal-to-noise ratio}
\newacro{IoU}{intersection-over-union}
\newacro{ICL}{in-context learning}

\usepackage{booktabs}

\usepackage{hyperref}

\usepackage{amsmath}
\usepackage{amssymb}
\usepackage{mathtools}
\usepackage{amsthm}

\theoremstyle{plain}

\theoremstyle{definition}

\theoremstyle{remark}

\usepackage{dcase2025,amsmath,graphicx,url,times,booktabs, tabularx}

\title{Synthetic data enables context-aware bioacoustic sound event detection}

\name{%
  \begin{tabular}{c}
  Benjamin Hoffman\textsuperscript{1},
  David Robinson\textsuperscript{1},
  Marius Miron\textsuperscript{1},
  Vittorio Baglione\textsuperscript{2},
  Daniela Canestrari\textsuperscript{2},
  Damian Elias\textsuperscript{3},\\
  \textit{Eva Trapote\textsuperscript{4},
  Felix Effenberger\textsuperscript{1},
  Maddie Cusimano\textsuperscript{1},
  Masato Hagiwara\textsuperscript{1},
  Olivier Pietquin\textsuperscript{1}}
  \end{tabular}
}

\address{%
\textsuperscript{1} Earth Species Project, USA\\
\textsuperscript{2} Universidad de Le\'on, Spain\\
\textsuperscript{3} University of California, Berkeley, USA\\
\textsuperscript{4} Universidad de Valladolid, Spain
}

\begin{document}

\maketitle

\begin{abstract}
We propose a methodology for training foundation models that enhances their in-context learning capabilities within the domain of bioacoustic signal processing. We use synthetically generated training data, introducing a domain-randomization-based pipeline that constructs diverse acoustic scenes with temporally strong labels. We generate over 8.8 thousand hours of strongly-labeled audio and train a query-by-example, transformer-based model to perform few-shot bioacoustic sound event detection. Our second contribution is a public benchmark of 13 diverse few-shot bioacoustics tasks. 
Our model outperforms previously published methods, and improves relative to other training-free methods by $64\%$. We demonstrate that this is due to increase in model size and data scale, as well as algorithmic improvements. We make our trained model available via an API, to provide ecologists and ethologists with a training-free tool for bioacoustic sound event detection.
\end{abstract}


\section{Introduction}
\label{introduction}

Foundation models can learn new tasks at inference time from a few labeled examples—a process known as few-shot or \ac{ICL}~\cite{Brown2020languagemodels}. This is attractive for application-driven ML domains—like bioacoustics, ecology, and conservation—where domain experts often lack ML experience and large labeled datasets \cite{robinson2025naturelm,miao2024newfrontiers}. Despite growing interest in adapting foundation models for these fields, data scarcity limits the tasks they can be trained to perform~\cite{miao2024newfrontiers}.

An example of this situation occurs in \ac{FSBSED}, which attempts to provide flexible modeling for the diversity of problems that arise in bioacoustics.
In this task, formalized in~\cite{NOLASCO2023Learning}, a model receives a \textit{support set}: 
an audio recording with onset and offset annotations for the first few events of interest. The model must predict onsets and offsets of these events in the \textit{query set}, which is the remainder of the recording. 

Temporally fine-scale detection is crucial for many applications in animal behavior and ecology~\cite{stowell2022computational}, but the time and expertise needed to annotate bioacoustic events has resulted in a lack of data available for training models capable of \ac{FSBSED}. Prior efforts for \ac{FSBSED} largely rely on a single 22-hour training dataset described in~\cite{NOLASCO2023Learning}, leading to lightweight models tailored to small-scale data.

\begin{figure}[htb]
\begin{center}
\centerline{
\includegraphics[width=\linewidth]{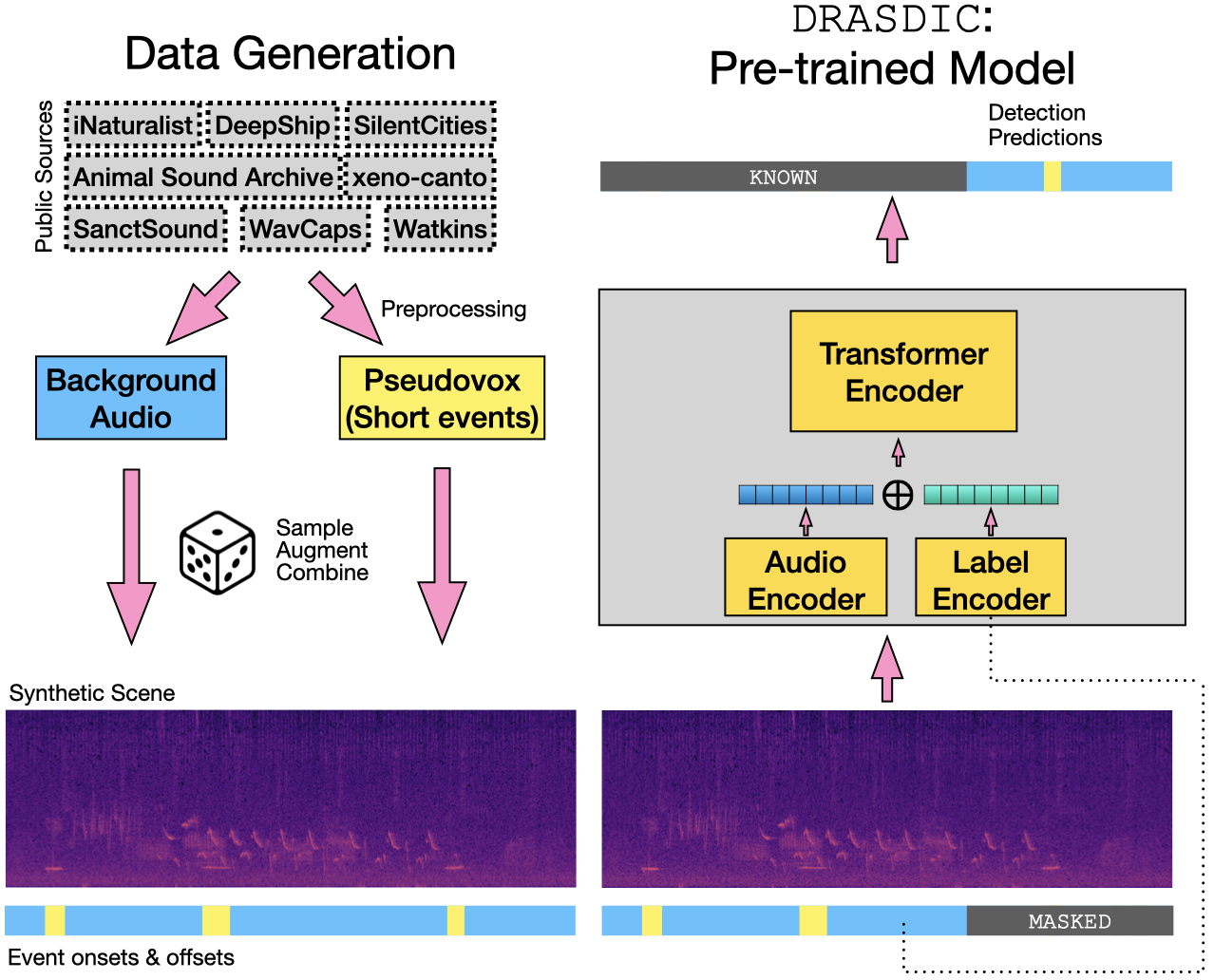}
}\vspace{-0.1in}\caption{
We introduce a method for generating synthetic acoustic scenes (Left) and a SotA few-shot detection model (Right).
}
\label{fig1}
\end{center}
\vspace{-1.5em}
\end{figure}

In this work, we investigate simultaneously scaling model parameters and training data volume, for a \ac{FSBSED} model tailored for \ac{ICL} (Figure~\ref{fig1}). To overcome limited annotated data, we turn to synthetic data, transforming raw audio into strongly labeled scenes via custom preprocessing and augmentations that introduce domain randomization. Because our focus is \ac{ICL}, we use a transformer-based few-shot model that attends to support and query audio jointly—common in \ac{ICL} but rare in \ac{FSBSED}. We call our method \method: \textit{Domain Randomization for Animal Sound Detection In-Context}.

To evaluate performance, we introduce a new 13-dataset \ac{FSBSED} benchmark, \benchmark (\textit{Fewshot Animal Sound Detection-13}).
\method achieves a $64\%$ average improvement over prior methods that also do not use gradient updates at inference. Ablations show improvements are due to simultaneously scaling model size, training data, and improving the few-shot mechanism.
We release \method weights, inference API, and \benchmark benchmark.\footnote{Available at \url{www.github.com/earthspecies/drasdic_api}.}

\section{Related Work}
\label{related_work}

\ac{FSBSED} was introduced in~\cite{NOLASCO2023Learning}. 
Challenges include sparse vocalizations, diverse target sounds, dynamic environments, and domain generalization~\cite{liang2024minddomaingapsystematic}.
Published methods include prototypical networks~\cite{surreysystem}, representation learning~\cite{Moummad2024contrastive}, and transductive inference~\cite{Yang2022Transductive}. Prior evaluation of \ac{FSBSED} systems has centered around the DCASE challenge~\cite{liang2024minddomaingapsystematic}, which provided public training and validation datasets and used a private test set. Subsequent efforts have either used the public validation set for both model selection and model evaluation~\cite{liang2024minddomaingapsystematic}, or skipped model selection~\cite{Moummad2024contrastive}. 

\Acf{ICL} refers to a model's ability to perform a task specified through demonstrations at inference time~\cite{Brown2020languagemodels}. 
\ac{ICL} has also been extended to fine-scale tasks in computer vision that somewhat resemble \ac{FSBSED}, which include sementic segmentation~\cite{bar2022visualpromptingimageinpainting,Wang_2023_CVPR} and scene understanding~\cite{Balazevic2023Scene}. Similar to our method,~\cite{Wang_2023_CVPR} employ a simple encoder-based architecture.

Generative vision and audio models have been used to create data for few-shot and low-resource tasks including detection~\cite{lin2023explorepowersyntheticdata}
We are not aware of a generative audio model that produces realistic and low-SNR animal sounds, and so instead developed a preprocessing pipeline to isolate potential animal sounds in publicly available data. A similar procedure was developed recently in~\cite{weldy2025simulated} for low-resource bioacoustic classification. For sound event detection in general audio, synthetic scenes assembled from multiple clips have been used to train models with fixed~\cite{Serizel2020SyntheticDomestic} and open ontologies~\cite{wu2025flam}.

\begin{figure}[htb]
\begin{center}
\includegraphics[width=\linewidth]{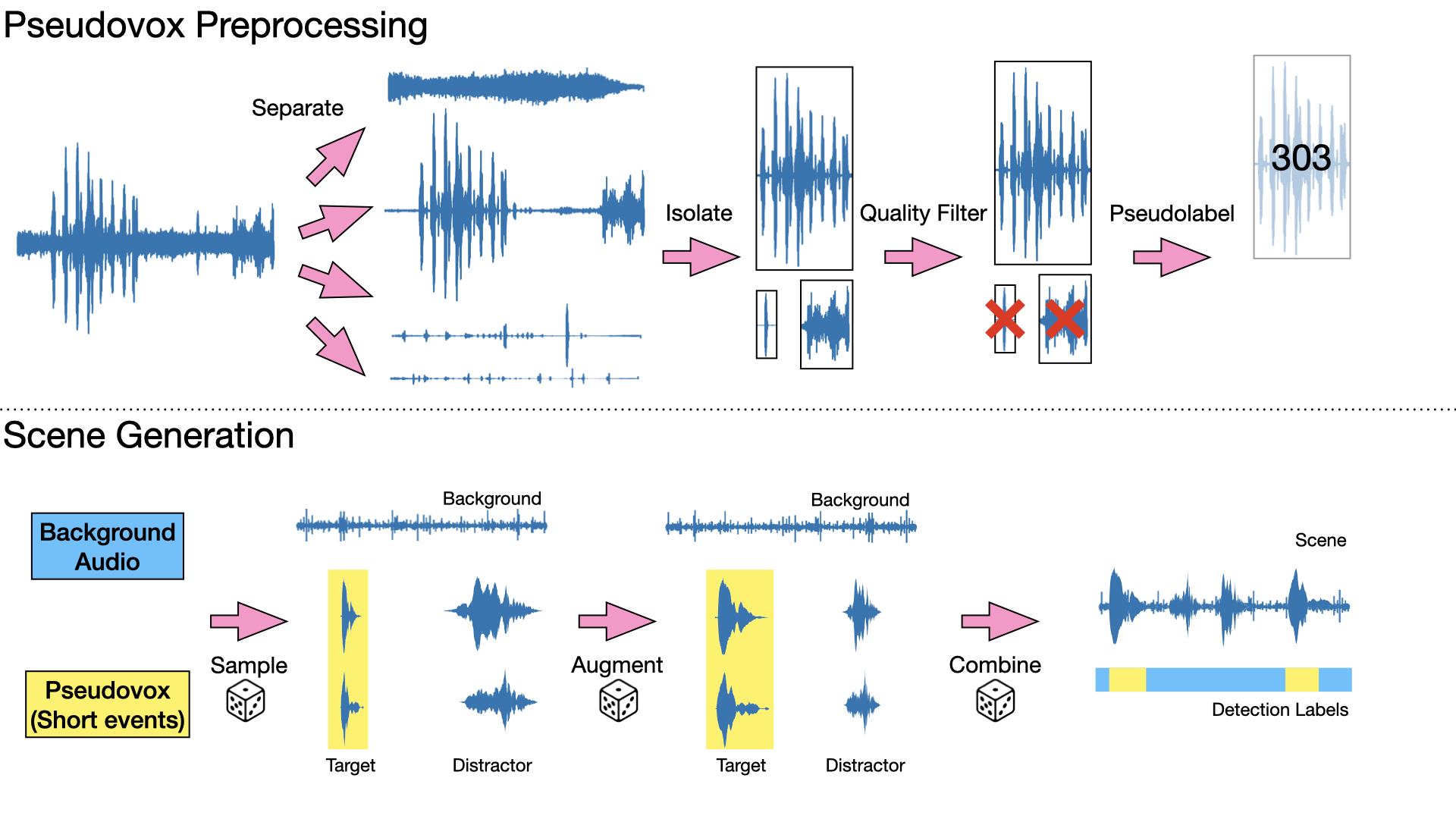}
\caption{Summary of training data preprocessing and scene generation.}
\label{data_generation}
\end{center}
\vspace{-1.5em}
\end{figure}

\section{Method}
\label{method}

\subsection{Data Generation}

We propose a two-stage approach to generate scenes (Figure~\ref{data_generation}). From publicly available unlabeled audio, we derive a set of background tracks (5.1e5 tracks, 5540 hours) and a set of short clips containing events dubbed pseudo-vocalizations (\textit{pseudovox}) (5.4e6 events, 577 hours); these are often animal vocalizations but may include non-biological acoustic events. These are pseudo-labeled through clustering, so multiple similar-sounding pseudovox can be sampled together.

In the second stage, performed on-the-fly during training, clips are randomly sampled from these collections, manipulated with data augmentations, and combined into scenes. Generated scenes may not always resemble real audio, due to  randomness in the scene generation process. As prior work has shown that domain randomization in synthetic data generation improves transfer to real data~\cite{Tobin2017DomainRF}, we view this as a way of increasing test-time robustness of our method. 

\subsubsection{Preprocessing} 
To construct our pseudovox set, we used public recordings from iNaturalist, Animal Sound Archive, xeno-canto\footnote{\url{www.inaturalist.org/}, \url{www.museumfuernaturkunde.berlin/en/research/animal-sound-archive}, \url{www.xeno-canto.org}, respectively}, as well as Watkins~\cite{sayigh2016watkins}, and WavCaps~\cite{mei2023wavcaps}. To remove background noise, we separated each recording into four stems using BirdMixIT~\cite{denton2022improving}. For each stem, we isolated potential pseudovox: segments where the amplitude envelope exceeded 25\% of the recording’s maximum, indicating a possible acoustic event. Many segments still lacked a clear acoustic event, so we performed a quality filtering step. We manually annotated a subset of segments for vocalization presence, then trained a binary linear classifier on the final layer BirdNET~\cite{kahl2021birdnet} activations for each segment. We applied this quality filter; passing segments became the final pseudovox set. This procedure resulted in $M=5.4e6$ pseudovox. Based on performance on a held-out test set, we estimated that 98\% of these pseudovox contained a clear acoustic event.
To obtain pseudolabels for the pseudovox, we applied $k$-means clustering to their BirdNET activations. We did this for $k\in K = \{\left \lfloor{M/128}\right \rfloor ,\left \lfloor{M/64}\right \rfloor,\left \lfloor{M/32}\right \rfloor,\left \lfloor{M/16}\right \rfloor,\left \lfloor{M/8}\right \rfloor\},$ to obtain different levels of cluster homogeneity. We inspected a random sample of 100 clusters, rating clusters as high- or low-quality based on acoustic homogeneity. Of these, 99 were deemed high quality. For background audio, we took the raw audio above, along with audio from
SilentCities~\cite{challeat2024acousticdataset}, DeepShip~\cite{irfan2021deepship}, and SanctSound~\cite{sanctsound}. 

\subsubsection{Scene Generation} Scene generation consists of three parts: sampling audio clips, manipulating them with data augmentations, and combining them to form a scene. In Section~\ref{ablationssection}, we investigate how the randomness in this process influences model performance.

We first sample two background tracks which are overlaid on each other. We choose a clustering level $k\in K$ and two clusters $c_{T}, c_{D}$ from the clusters of level $k$. We sample a random number of target pseudovox from $c_{T}$, and a random number of distractor pseudovox from $c_{D}$. We apply reverb (drawn from~\cite{rir_data}), resampling, time flipping, and amplitude augmentations to pseudovox, and resampling augmentations to background tracks. We paste pseudovox into the background track, one-by-one, with a random time gap between pseudovox. 
We maintain a binary annotation mask for the scene. This mask is initialized with zeros, and changed to ones where target pseudovox are added. Distractor pseudovox do not change the mask; they join whatever sounds are already present in the background tracks. To generate one training example, two scenes (support and query) are generated, drawing pseudovox from the same $c_T,c_D$ for both. With some probability, the background tracks of the query are chosen to be different than those of the support.

\subsection{Model}

Using our synthetic scenes, we train our model \method. During training the model is given annotated support audio and unannotated query audio, and must predict detection labels for the query audio. 

\subsubsection{Architecture} Noting that encoder-only architectures have been used successfully for fine-scale \ac{ICL} problems in computer vision~\cite{Wang_2023_CVPR}, we adopt a simple but highly parametrized BERT-like architecture which applies attention to support and query simultaneously. This is preceded by a CNN spectrogram encoder. 

Support and query audio are resampled to 16 kHz, concatenated, and converted to a log mel-spectrogram (256 mels, hop size 160). The CNN encoder is a 2-d convolutional block and two 2-d residual blocks (ker=7, 3, 3, respectively; hidden size 64), with vertical mean pooling (ker=2) after each. Frequency and hidden dimensions are flattened and mean-pooled to a final 50 Hz frame rate. The binary support label mask is max-pooled to 50 Hz, passed to a per-frame label embedding, and added to the encoded audio. This label-enriched representation enters a transformer encoder (hidden size 768, 12 heads, 12 blocks) with rotary position encoding~\cite{SU2024Rope}.
A final linear layer maps each frame to detection logits.

\subsubsection{Training} \label{sstraining} \method was randomly initialized and trained with per-frame binary cross-entropy loss on the query labels, using AdamW~\cite{loshchilov2018decoupled} with $(\beta_0,\beta_1)=(0.9, 0.999)$ and weight decay $0.01$. We used support-query pairs of total duration $\operatorname{dur}_s+\operatorname{dur}_q$ seconds. Based on initial experiments, we set $\operatorname{dur}_s=30$ and $\operatorname{dur}_q=10$.

Model, data generation, and training hyperparameters were chosen through random search.  As our model selection criterion, we used average performance on the validation datasets from~\cite{liang2024minddomaingapsystematic}.
We applied curriculum learning to gradually increase task difficulty during training. This linearly decays the minimum pseudovox \ac{SNR} from 0 dB to a minimum of -20 dB for an initial $5e4$ steps. The learning rate is linearly increased for $1e4$ steps to a maximum of $2e\!-\!5$, and then decayed to $0$ after $1e5$ steps (cosine schedule) using batch size of 8.
Parameters governing data generation are provided in the GitHub repository. 

\section{Public Benchmark}
\label{dataset}

\begin{table*}[hbt]
\caption{Details of \benchmark. Datasets were chosen for their taxonomic diversity, varied recording conditions, and quality of their annotations. They were manually subsetted (prior to evaluation), to reduce computational overhead. Other (minor) preprocessing steps are described on the project GitHub. Datasets with a $\dagger$ are presented for the first time here. Terrestrial and underwater autonomous passive acoustic monitoring devices are abbreviated T. PAM and U. PAM, respectively.}
\label{dataset_summary}
\begin{center}
\begin{scriptsize}
\resizebox{\textwidth}{!}{%
\begin{tabular}{llccccccc}
\toprule
Dataset & Full Name & N files & Dur (hr) & N events & Recording type & Location & Taxa & Detection target \\
\midrule
AS~\cite{canas2023dataset} & AnuraSet & 12 & 0.20 & 162 & T. PAM & Brazil & Anura & Species \\
CC$^\dagger$ & Carrion Crow & 10 & 10.00 & 2200 & On-body & Spain & \makecell{Corvus corone + \\Clamator glandarius} & \makecell{Species +  \\ Life Stage} \\
GS~\cite{YOH2024gunshots} & Gunshot & 7 & 38.33 & 85 & T. PAM & Gabon & Homo sapiens & Production Mechanism \\
HA~\cite{navine2022soundscape} & Hawaiian Birds & 12 & 1.10 & 628 & T. PAM & Hawaii, USA & Aves & Species \\
HG~\cite{dufourq2021automated} & Hainan Gibbon & 9 & 72.00 & 483 & T. PAM & Hainan, China & Nomascus hainanus & Species \\
HW~\cite{Allen2021} & Humpback Whale & 10 & 2.79 & 1565 & U. PAM & North Pacific Ocean & Megaptera novaeangliae & Species \\
JS$^\dagger$ & Jumping Spider & 4 & 0.23 & 924 & Substrate & Laboratory & Habronattus & Sound Type \\
KD~\cite{symes2023neotropical} & Katydid & 12 & 2.00 & 883 & T. PAM & Panam\'a & Tettigoniidae & Species \\
MS~\cite{sarkar2023can,zhang2018automatic} & Marmoset & 10 & 1.67 & 1369 & Laboratory & Laboratory & Callithrix jacchus & Vocalization Type \\
PM~\cite{Chronister2021Powdermill} & Powdermill & 4 & 6.42 & 2032 & T. PAM & Pennsylvania, USA & Passeriformes & Species \\
RG~\cite{Lapp2023rg} & Ruffed Grouse & 2 & 1.50 & 34 & T. PAM & Pennsylvania, USA & Bonasa umbellus & Species \\
RS~\cite{Lapp2024rana} & Rana Sierrae & 7 & 1.87 & 552 & U. PAM & California, USA & Rana sierrae & Species \\
RW~\cite{kirsebom2020performance} & Right Whale & 10 & 5.00 & 398 & U. PAM & Gulf of St. Lawrence & Eubalaena glacialis & Species \\
\bottomrule
\end{tabular}
}
\end{scriptsize}
\end{center}
\vspace{-1.5em}
\end{table*}

A collection of public \ac{FSBSED} datasets was previously provided in~\cite{NOLASCO2023Learning,liang2024minddomaingapsystematic}, but were designated as datasets for model training and validation. We complement these with \benchmark, a public benchmark curated for model evaluation (Table~\ref{dataset_summary}). \benchmark consists of 13 bioacoustics datasets, each of which includes between 2 and 12 audio files. Eleven of these datasets were derived from previous studies; they were chosen for their taxonomic diversity, varied recording conditions, and quality of their annotations. Two (CC and JS) are presented here for the first time. All datasets were developed alongside studies of ecology or animal behavior, and represent a range of realistic problems encountered in bioacoustics data. Details of dataset collection and preprocessing steps are available at the GitHub repository.

We follow the data format in~\cite{NOLASCO2023Learning}: Each audio file comes with annotations of the onsets and offsets of \textit{positive} sound events, i.e.~sounds coming from a predetermined category (such as a species label or call type). 
An $N$-shot detection system is provided with the audio up through the $N^{th}$ positive event, and must predict the onsets and offsets of positive events in the rest of the recording.

\section{Experimental Evaluation}
\label{results}

\begin{table*}[t]
\caption{F1 scores @0.3 IoU on \benchmark. Methods marked with $\dagger$ were pre-trained using our generated data, rather than the data used in the original publication. Methods marked with $^*$ involve no gradient updates at inference time. The second column gives the number of model parameters, and the final column gives the average F1 score across the six validation datasets from~\cite{liang2024minddomaingapsystematic}.}
\label{main_results}
\begin{center}
\begin{small}

\vspace{-1.5em}
\begin{tabular}{l|l|ccccccccccccc|c|c}
\toprule


\midrule
Model & Params & AS & CC & GS & HA & HG & HW & JS & KD & MS & PM & RG & RS & RW & Avg & Val \\
\midrule

BEATS+linear & 90M & .350 & .003 & .056 & .093 & \textbf{.242} & .173 & .028 & .049 & .462 & .212 & \textbf{.732} & .007 & .316 &.209 & .358 \\

AVES+linear & 90M &.586 &.059 &.500 &.374 &.207 &.366 &.026 &\textbf{.673} &\textbf{.831} &.291 &.529 &.303 &.494 &.403  & .565 \\

Protonet$^*$ & 0.7M & .356 & .189 & .156 & .239 & .038 & .085 & .136 & .316 & .590 & .260 & .000 & .216 & .393 &.229 & .461 \\

Protonet$^{\dagger *}$  & 0.7M &.305 &.224 &.151 &.307 &.023 &.116 &.166 &.418 &.536 &.235 &.121 &.195 &.342 &.242 & .459 \\

Transductive  & 0.5M &.299 &.144 &.002 &.283 &.020 &.116 &.279 &.218 &.569 &.159 &.089 &.169 &.048 &.184  & .242 \\

SCL$^*$  & 7.2M & .516 & .333 & .025 & .438 & .010 & .255 & .281 & .263 & .402 & .237 & .049 & .219 & .509 &.272 & .514\\

SCL+finetuning  & 7.2M & .565 & \textbf{.341} & .017 & .467 & .008 & .382 & \textbf{.302} & .381 & .476 & .327 & .042 & .285 & .275 & .298 & .525\\

SCL$^{\dagger*}$  & 7.2M  &.545 &.287 & .024 &.433 &.008 &.393 & .243 &.207 & .429 & .336 & .038 &.218 &.228 &.261 & .440\\

SCL$^\dagger$+finetuning  & 7.2M & .571 & .205 & .030 & .479 & .005 & \textbf{.453} & .132 & .220 & .516 & .450 &.050 &.292 &.223 &.279  &.453\\

\method$^{\dagger*}$ (ours)  & 116M &\textbf{.645} &.272 &\textbf{.593} &\textbf{.587} &.144 &.337 &.099 &.644 &.783 &\textbf{.474} &.092 &\textbf{.352} &\textbf{.764} &\textbf{.445}  & \textbf{.704}\\

\bottomrule

\end{tabular}
\end{small}
\end{center}
\vspace{-1.5em}
\end{table*}

We evaluate models based on their ability to detect events after the $N=5^{th}$ positive event in each recording of \benchmark, using F1@0.3 IoU as described in~\cite{NOLASCO2023Learning}.
We used performance on the validation set from~\cite{liang2024minddomaingapsystematic} to select a final model to evaluate on \benchmark.

\subsection{Inference} For \method, we form predictions by windowing the audio in each recording, making multiple predictions for each window by prompting the model multiple times, and averaging these predictions. In detail, for a fixed $\operatorname{dur}_q$-second window of the query set, we prompt the model $N=5$ times and average the frame-wise predictions produced by these five prompts. The support set for the $i^{th}$ prompt ($i\in[1,5]$) is the $\operatorname{dur}_s$-seconds of support audio centered at the $i^{th}$ positive event in the support set (together with the binary detection mask). This procedure is repeated for $\operatorname{dur}_q$-second windows across the entire query set. Frames with predicted detection probability above a fixed threshold of 0.5 become positive detections. These are smoothed: detections separated by a gap of $\min(1,d/2)$ seconds are merged, and then detections lasting less than $\min(1/2,d/2)$ seconds are discarded.  Here $d$ is the duration of the shortest event in the support set. 

\subsection{Comparison methods}

We compare \method with essentially all of the previously published methods we are aware of for \ac{FSBSED} that contain publicly available implementations. The first, ``BEATS+linear'' is a simple supervised baseline which consists of a frozen BEATS encoder~\cite{Chen2023Beats} and a final linear layer. Support audio is windowed (4 seconds, 50\% overlap) and the final layer is trained for 100 epochs to predict binary per-frame detection labels (final frame rate: 50 Hz). Training minimizes average per-frame binary cross-entropy loss.
The initial learning rate of 0.01 (tuned using the validation set) is decayed to 0 using a cosine schedule. 
The second ``AVES+linear'' replaces the BEATS encoder with the pre-trained AVES encoder~\cite{hagiwara2023aves} (BirdAVES Base checkpoint), which was pre-trained on 2570 hours of animal sounds.
``Protonet'' is the prototypical network from~\cite{liang2024minddomaingapsystematic}, which itself adapts~\cite{surreysystem}. ``Transductive''~\cite{Yang2022Transductive} uses a CNN encoder that is updated using unlabeled audio from the query set. ``SCL'' applies the supervised contrastive learning method introduced by~\cite{Moummad2024contrastive}. ``SCL+finetuning'', also introduced by~\cite{Moummad2024contrastive} extends this by using support audio to fine-tune the encoder that was pre-trained using the SCL method.

For Protonet, SCL, and SCL+Finetuning, we train a version using the training data from~\cite{NOLASCO2023Learning}. We also train a version using our generated scenes ($5e4$ scenes, each 40 seconds), which represents a $ 25\times$ increase in data quantity over the data used for training the original models. 

\begin{figure}[htb]
\begin{center}
\includegraphics[width=\linewidth]{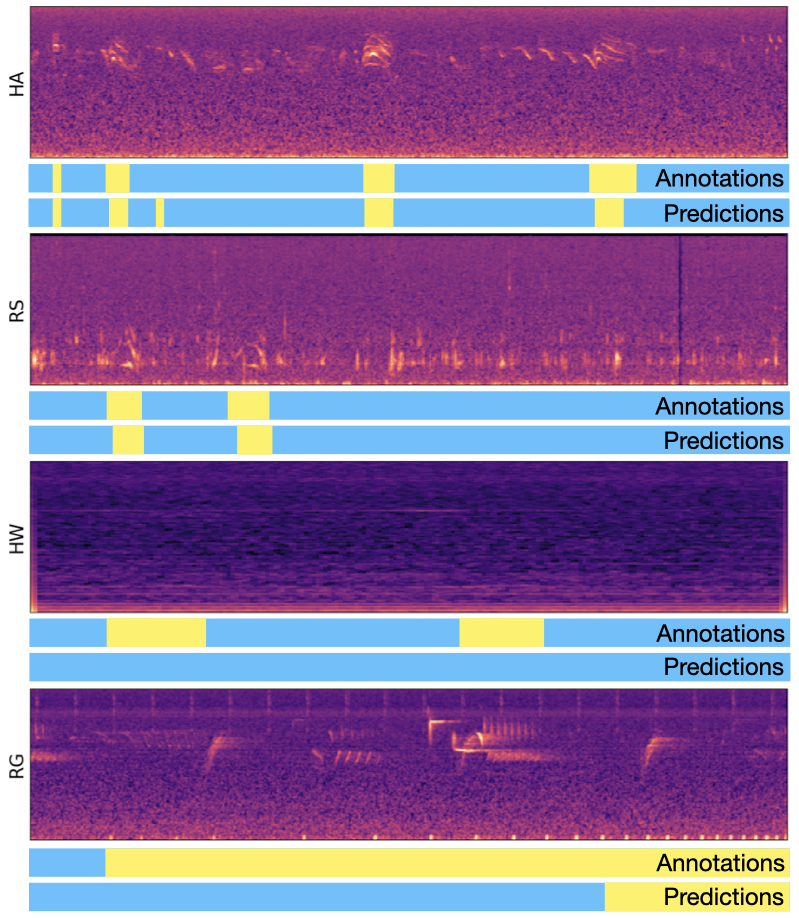}
\vspace{-2em}
\caption{Qualitative results; events are in yellow. \method detects target sounds in dynamic environments (top two), but challenges include extremely low \ac{SNR} (third), and the extended low-frequency drumming displays of ruffed grouse (bottom). Each spectrogram represents 10 seconds of audio.}
\label{exampleresults}
\end{center}
\vspace{-1.5em}
\end{figure}

\subsection{Experiments}

We compare model performance on \benchmark (Table~\ref{main_results}). Some datasets (JS, KD, MS, PB, PB24) contain events that are above \method's 8kHz Nyquist frequency, or that are brief relative to the model's 50Hz frame rate. For these, we give the model a 1/2-time version (1/6 for KD). We give other methods \textit{both} the slowed and full-speed version of the data, and keep the version with the better score.

On \benchmark, \method outperforms all the alternatives on 6 out of 13 datasets. Across datasets \method has an average F1 score of .042 over the next best model. Compared to other methods that do not require gradient updates at inference time, \method outperforms the others on 9 of 13 datasets, and has an average F1 score of .173 over the next best model (64\% relative improvement).
Qualitatively, \method detected diverse target sounds, even amid others in the same frequency bands (Figure~\ref{exampleresults}, top). Performance is strong across a variety of taxa and conditions. A failure case is for the JS dataset, which consists of jumping spider drumming. Here, the detection targets are specific drum types, and distinguishing between drum types relies partly on the rate of drumming. Our scene generator did not account for this type of information. Other failure cases are in Figure~\ref{exampleresults}, bottom.

For the comparison methods we trained with our generated data, there was no clear performance increase. These methods, which adopt a CNN architecture, employ a different few-shot mechanism than \method and also have fewer than $1/10$ the trainable parameters. The relative impact of these differences is investigated in Section~\ref{ablationssection}.

\subsection{Ablation experiments} \label{ablationssection}

In our main experiments, we scaled model parameters and data volume, while also adopting a few-shot mechanism that applied attention to support and query audio simultaneously. We conducted experiments to investigate the contributions of these changes, individually (Table~\ref{ablations_table}, top). First, we compared our main model (116.1M parameters), whose transformer encoder has the same structure as BERT Base, to smaller versions based on BERT Small~\cite{DBLP:journals/corr/abs-1908-08962} (19.3M parameters) and BERT Tiny (2.5M parameters). Second, we compared our main data generation procedure to one that only generated 220 hours of unique scenes, and one that only generated 22 hours of unique scenes. Additionally, we compared to a version that used the non-synthetic training data (22 hours total) from~\cite{NOLASCO2023Learning}, as well as a version for which 10\% of the training examples were from~\cite{NOLASCO2023Learning} and the other 90\% synthetic. Finally, we adjusted our few-shot mechanism to a prototypical network, which prevented attention from being applied to support and query audio simultaneously. For this, we kept the same architecture as our main method but applied a prototypical loss as in~\cite{liang2024minddomaingapsystematic, surreysystem}. Average performance on \benchmark and on the validation set dropped in all cases, indicating that each of these changes contributed to final model performance. Reducing data scale was especially damaging, likely due to the high number of trainable parameters in our main method.

We investigated the 
impacts of adjusting the randomness governing our scene generation procedure (Table~\ref{ablations_table}, bottom). We perturbed the level of homogeneity of target events in a scene, the typical rate of events, the loudness of events, and whether we apply pitch shifting augmentations.
Average performance is stable across some of these perturbations, but decreases when the randomness in event rate and event SNR is decreased. These parameters likely influence the level of diversity present across generated scenes more than the others.

Eliminating random pitch shifts resulted in slightly better performance on \benchmark. Designing a domain randomization strategy is an optimization problem, which we approached through a model selection criterion. This criterion did not produce the best model on the test set, aligning with the observation \cite{liang2024minddomaingapsystematic} that strong domain shifts between few-shot tasks present a challenge for \ac{FSBSED} model development. 

\begin{table}[t]
\scriptsize
\caption{Average F1 scores @0.3 IoU on \benchmark and validation~\cite{liang2024minddomaingapsystematic} datasets. Model ablations appear on top, data ablations on bottom.}

\vspace{-1.5em}
\label{ablations_table}
\begin{center}
\resizebox{\linewidth}{!}{%
\begin{tabular}{l|cc}
\toprule
Method  & Avg (test)  & Avg (val) \\
\midrule
\method  &.445  &.704  \\
\midrule
BERT Small &.425 &.666  \\
BERT Tiny &.323  &.521  \\
Reduced data (220 h) &.354 &.504  \\
Reduced data (22 h)  &.130  &.088  \\
Non-synthetic data (22 h) &.135  &.165  \\
10\% non-synthetic data &.428 &.666  \\
Protonet loss &.444 &.638  \\
\midrule
High homogeneity in events  &.429 &.630  \\
Low homogeneity in events  &.444  &.606  \\
High events / second   &.381 &.437  \\
Low events / second  &.370 &.669  \\
Only high SNR events  &.402 &.620  \\
Only low SNR events  &.438 &.682  \\
No pitch/time shifting  &.457 &.613  \\
\bottomrule
\end{tabular}
}
\end{center}
\vspace{-1.5em}
\end{table}

\section*{Conclusion}

To provide a training-free solution for fine-scale bioacoustic sound event detection, we develop a \ac{ICL} transformer model \method. We develop a domain-randomization based data-generation pipeline, and train our model on over 8.8 thousand hours of synthetic acoustic scenes. We additionally provide \benchmark, a new benchmark  for few-shot bioacoustic sound event detection.
Our model substantially improves upon previous state-of-the art. We demonstrate that these improvements are due to both our modeling approach and the data scale provided by our scene generation method.

\bibliographystyle{IEEEtran}
\bibliography{refs}

\begin{thebibliography}{10}
\providecommand{\url}[1]{#1}
\csname url@samestyle\endcsname
\providecommand{\newblock}{\relax}
\providecommand{\bibinfo}[2]{#2}
\providecommand{\BIBentrySTDinterwordspacing}{\spaceskip=0pt\relax}
\providecommand{\BIBentryALTinterwordstretchfactor}{4}
\providecommand{\BIBentryALTinterwordspacing}{\spaceskip=\fontdimen2\font plus
\BIBentryALTinterwordstretchfactor\fontdimen3\font minus \fontdimen4\font\relax}
\providecommand{\BIBforeignlanguage}[2]{{%
\expandafter\ifx\csname l@#1\endcsname\relax
\typeout{** WARNING: IEEEtran.bst: No hyphenation pattern has been}%
\typeout{** loaded for the language `#1'. Using the pattern for}%
\typeout{** the default language instead.}%
\else
\language=\csname l@#1\endcsname
\fi
#2}}
\providecommand{\BIBdecl}{\relax}
\BIBdecl

\bibitem{Brown2020languagemodels}
T.~Brown, B.~Mann, N.~Ryder, M.~Subbiah, J.~D. Kaplan \emph{et~al.}, ``Language models are few-shot learners,'' in \emph{Advances in Neural Information Processing Systems}, H.~Larochelle, M.~Ranzato, R.~Hadsell, M.~Balcan, and H.~Lin, Eds., vol.~33, 2020, pp. 1877--1901.

\bibitem{robinson2025naturelm}
D.~Robinson, M.~Miron, M.~Hagiwara, and O.~Pietquin, ``{NatureLM}-audio: An audio-language foundation model for bioacoustics,'' in \emph{Proceedings of the International Conference on Learning Representations (ICML)}, 2025.

\bibitem{miao2024newfrontiers}
\BIBentryALTinterwordspacing
Z.~Miao, Y.~Zhang, Z.~Fabian, A.~Hernandez~Celis, S.~Beery \emph{et~al.}, ``\BIBforeignlanguage{English}{{New frontiers in AI for biodiversity research and conservation with multimodal language models}},'' \emph{\BIBforeignlanguage{English}{Preprint}}, 2024. [Online]. Available: \url{https://doi.org/10.32942/X22S6F}
\BIBentrySTDinterwordspacing

\bibitem{NOLASCO2023Learning}
I.~Nolasco, S.~Singh, V.~Morfi, V.~Lostanlen, A.~Strandburg-Peshkin \emph{et~al.}, ``Learning to detect an animal sound from five examples,'' \emph{Ecological Informatics}, vol.~77, p. 102258, 2023.

\bibitem{stowell2022computational}
D.~Stowell, ``Computational bioacoustics with deep learning: a review and roadmap,'' \emph{PeerJ}, vol.~10, p. e13152, 2022.

\bibitem{liang2024minddomaingapsystematic}
J.~Liang, I.~Nolasco, B.~Ghani, H.~Phan, E.~Benetos, and D.~Stowell, ``Mind the domain gap: A systematic analysis on bioacoustic sound event detection,'' in \emph{2024 32nd European Signal Processing Conference (EUSIPCO)}, 2024, pp. 1257--1261.

\bibitem{surreysystem}
H.~Liu, X.~Liu, X.~Mei, Q.~Kong, W.~Wang, and M.~D. Plumbley, ``Segment-level metric learning for few-shot bioacoustic event detection,'' \emph{DCASE}, 2022.

\bibitem{Moummad2024contrastive}
I.~Moummad, N.~Farrugia, and R.~Serizel, ``Regularized contrastive pre-training for few-shot bioacoustic sound detection,'' in \emph{ICASSP 2024 - 2024 IEEE International Conference on Acoustics, Speech and Signal Processing (ICASSP)}, 2024, pp. 1436--1440.

\bibitem{Yang2022Transductive}
D.~Yang, H.~Wang, Y.~Zou, Z.~Ye, and W.~Wang, ``A mutual learning framework for few-shot sound event detection,'' in \emph{ICASSP 2022 - 2022 IEEE International Conference on Acoustics, Speech and Signal Processing (ICASSP)}, 2022, pp. 811--815.

\bibitem{bar2022visualpromptingimageinpainting}
A.~Bar, Y.~Gandelsman, T.~Darrell, A.~Globerson, and A.~Efros, ``Visual prompting via image inpainting,'' \emph{Advances in Neural Information Processing Systems}, vol.~35, pp. 25\,005--25\,017, 2022.

\bibitem{Wang_2023_CVPR}
X.~Wang, W.~Wang, Y.~Cao, C.~Shen, and T.~Huang, ``Images speak in images: A generalist painter for in-context visual learning,'' in \emph{Proceedings of the IEEE/CVF Conference on Computer Vision and Pattern Recognition (CVPR)}, June 2023, pp. 6830--6839.

\bibitem{Balazevic2023Scene}
I.~Balazevic, D.~Steiner, N.~Parthasarathy, R.~Arandjelovi\'{c}, and O.~Henaff, ``Towards in-context scene understanding,'' in \emph{Advances in Neural Information Processing Systems}, vol.~36, 2023, pp. 63\,758--63\,778.

\bibitem{lin2023explorepowersyntheticdata}
S.~Lin, K.~Wang, X.~Zeng, and R.~Zhao, ``Explore the power of synthetic data on few-shot object detection,'' in \emph{Proceedings of the IEEE/CVF conference on computer vision and pattern recognition}, 2023, pp. 638--647.

\bibitem{weldy2025simulated}
M.~J. Weldy, D.~B. Lesmeister, T.~Denton, A.~Duarte, B.~J. Vernasco \emph{et~al.}, ``Simulated soundscapes and transfer learning boost the performance of acoustic classifiers under data scarcity,'' \emph{Methods in Ecology and Evolution}, 2025.

\bibitem{Serizel2020SyntheticDomestic}
R.~Serizel, N.~Turpault, A.~Shah, and J.~Salamon, ``Sound event detection in synthetic domestic environments,'' in \emph{ICASSP 2020 - 2020 IEEE International Conference on Acoustics, Speech and Signal Processing (ICASSP)}, 2020, pp. 86--90.

\bibitem{wu2025flam}
Y.~Wu, C.~Tsirigotis, K.~Chen, C.-Z.~A. Huang, A.~Courville \emph{et~al.}, ``{FLAM}: Frame-wise language-audio modeling,'' in \emph{Forty-second International Conference on Machine Learning}, 2025.

\bibitem{Tobin2017DomainRF}
J.~Tobin, R.~Fong, A.~Ray, J.~Schneider, W.~Zaremba, and P.~Abbeel, ``Domain randomization for transferring deep neural networks from simulation to the real world,'' \emph{2017 IEEE/RSJ International Conference on Intelligent Robots and Systems (IROS)}, pp. 23--30, 2017.

\bibitem{sayigh2016watkins}
L.~Sayigh, M.~A. Daher, J.~Allen, H.~Gordon, K.~Joyce \emph{et~al.}, ``The {W}atkins marine mammal sound database: An online, freely accessible resource,'' \emph{Proceedings of Meetings on Acoustics}, vol.~27, no.~1, p. 040013, 2016.

\bibitem{mei2023wavcaps}
X.~Mei, C.~Meng, H.~Liu, Q.~Kong, T.~Ko \emph{et~al.}, ``Wav{C}aps: A {ChatGPT}-assisted weakly-labelled audio captioning dataset for audio-language multimodal research,'' \emph{IEEE/ACM Transactions on Audio, Speech, and Language Processing}, pp. 3339--3354, 2024.

\bibitem{denton2022improving}
T.~Denton, S.~Wisdom, and J.~R. Hershey, ``Improving bird classification with unsupervised sound separation,'' in \emph{ICASSP 2022-2022 IEEE International Conference on Acoustics, Speech and Signal Processing (ICASSP)}.\hskip 1em plus 0.5em minus 0.4em\relax IEEE, 2022, pp. 636--640.

\bibitem{kahl2021birdnet}
S.~Kahl, C.~M. Wood, M.~Eibl, and H.~Klinck, ``{BirdNET: A deep learning solution for avian diversity monitoring},'' \emph{Ecological Informatics}, vol.~61, p. 101236, 2021.

\bibitem{challeat2024acousticdataset}
S.~Challéat, N.~Farrugia, J.-S.~P. Froidevaux \emph{et~al.}, ``A dataset of acoustic measurements from soundscapes collected worldwide during the covid-19 pandemic,'' \emph{Scientific Data}, vol.~11, p. 928, 2024.

\bibitem{irfan2021deepship}
M.~Irfan, Z.~Jiangbin, S.~Ali, M.~Iqbal, Z.~Masood, and U.~Hamid, ``Deepship: An underwater acoustic benchmark dataset and a separable convolution based autoencoder for classification,'' \emph{Expert Systems with Applications}, vol. 183, p. 115270, 2021.

\bibitem{sanctsound}
NOAA, ``Passive acoustic data collection,'' 2017.

\bibitem{rir_data}
J.~Traer and J.~H. McDermott, ``Statistics of natural reverberation enable perceptual separation of sound and space,'' \emph{Proceedings of the National Academy of Sciences}, vol. 113, no.~48, pp. E7856--E7865, 2016.

\bibitem{SU2024Rope}
J.~Su, M.~Ahmed, Y.~Lu, S.~Pan, W.~Bo, and Y.~Liu, ``Roformer: Enhanced transformer with rotary position embedding,'' \emph{Neurocomputing}, vol. 568, p. 127063, 2024.

\bibitem{loshchilov2018decoupled}
I.~Loshchilov and F.~Hutter, ``Decoupled weight decay regularization,'' in \emph{International Conference on Learning Representations}, 2019.

\bibitem{canas2023dataset}
J.~S. Cañas, M.~P. Toro-Gómez, L.~S.~M. Sugai \emph{et~al.}, ``A dataset for benchmarking neotropical anuran calls identification in passive acoustic monitoring,'' \emph{Scientific Data}, vol.~10, p. 771, 2023.

\bibitem{YOH2024gunshots}
N.~Yoh, W.~Mbamy, B.~L. Gottesman, G.~Z. Froese, T.~Satchivi \emph{et~al.}, ``Impacts of logging, hunting, and conservation on vocalizing biodiversity in gabon,'' \emph{Biological Conservation}, vol. 296, p. 110726, 2024.

\bibitem{navine2022soundscape}
\BIBentryALTinterwordspacing
A.~Navine, S.~Kahl, A.~Tanimoto-Johnson, H.~Klinck, and P.~Hart, ``A collection of fully-annotated soundscape recordings from the island of {H}awai'i,'' 2022. [Online]. Available: \url{https://doi.org/10.5281/zenodo.7078499}
\BIBentrySTDinterwordspacing

\bibitem{dufourq2021automated}
E.~Dufourq, I.~Durbach, J.~P. Hansford, A.~Hoepfner, H.~Ma \emph{et~al.}, ``Automated detection of {H}ainan gibbon calls for passive acoustic monitoring,'' \emph{Remote Sensing in Ecology and Conservation}, vol.~7, no.~3, pp. 475--487, 2021.

\bibitem{Allen2021}
A.~N. Allen, M.~Harvey, L.~Harrell, A.~Jansen, K.~P. Merkens \emph{et~al.}, ``A convolutional neural network for automated detection of humpback whale song,'' \emph{Frontiers in Marine Science}, vol.~8, p. 607321, 2021.

\bibitem{symes2023neotropical}
\BIBentryALTinterwordspacing
L.~Symes, S.~Madhusudhana, S.~Martinson \emph{et~al.}, ``Neotropical forest soundscapes with call identifications for katydids,'' Dataset, 2023. [Online]. Available: \url{https://datadryad.org/dataset/doi:10.5061/dryad.zw3r2288b}
\BIBentrySTDinterwordspacing

\bibitem{sarkar2023can}
E.~Sarkar and M.~Magimai.-Doss, ``Can self-supervised neural representations pre-trained on human speech distinguish animal callers?'' in \emph{Proceedings of INTERSPEECH 2023}, 2023, pp. 1189--1193.

\bibitem{zhang2018automatic}
Y.-J. Zhang, J.-F. Huang, N.~Gong, Z.-H. Ling, and Y.~Hu, ``Automatic detection and classification of marmoset vocalizations using deep and recurrent neural networks,'' \emph{The Journal of the Acoustical Society of America}, vol. 144, no.~1, p. 478, 2018.

\bibitem{Chronister2021Powdermill}
L.~M. Chronister, T.~A. Rhinehart, A.~Place, and J.~Kitzes, ``An annotated set of audio recordings of eastern north american birds containing frequency, time, and species information,'' \emph{Ecology}, vol. 102, no.~6, p. e03329, 2021.

\bibitem{Lapp2023rg}
S.~Lapp, J.~L. Larkin, H.~A. Parker, J.~T. Larkin, D.~R. Shaffer \emph{et~al.}, ``Automated recognition of ruffed grouse drumming in field recordings,'' \emph{Wildlife Society Bulletin}, vol.~47, no.~1, p. e1395, 2023.

\bibitem{Lapp2024rana}
S.~Lapp, T.~C. Smith, R.~A. Knapp, A.~Lindauer, and J.~Kitzes, ``Aquatic soundscape recordings reveal diverse vocalizations and nocturnal activity of an endangered frog,'' \emph{The American Naturalist}, vol. 203, no.~5, pp. 618--627, 2024, pMID: 38635364.

\bibitem{kirsebom2020performance}
O.~S. Kirsebom, F.~Frazao, Y.~Simard, N.~Roy, S.~Matwin, and S.~Giard, ``Performance of a deep neural network at detecting north atlantic right whale upcalls,'' \emph{The Journal of the Acoustical Society of America}, vol. 147, no.~4, pp. 2636--2646, 2020.

\bibitem{Chen2023Beats}
S.~Chen, Y.~Wu, C.~Wang, S.~Liu, D.~Tompkins \emph{et~al.}, ``{BEAT}s: audio pre-training with acoustic tokenizers,'' in \emph{Proceedings of the 40th International Conference on Machine Learning}, ser. ICML'23, 2023.

\bibitem{hagiwara2023aves}
M.~Hagiwara, ``{AVES}: Animal vocalization encoder based on self-supervision,'' in \emph{ICASSP 2023-2023 IEEE International Conference on Acoustics, Speech and Signal Processing (ICASSP)}.\hskip 1em plus 0.5em minus 0.4em\relax IEEE, 2023, pp. 1--5.

\bibitem{DBLP:journals/corr/abs-1908-08962}
I.~Turc, M.~Chang, K.~Lee, and K.~Toutanova, ``Well-read students learn better: The impact of student initialization on knowledge distillation,'' \emph{CoRR}, vol. abs/1908.08962, 2019.

\end{thebibliography}


\newpage
\onecolumn

\appendix

\subsection{\benchmark Summary}

For all datasets, if there were overlapping bounding boxes corresponding to multiple events, we merged these bounding boxes into a single bounding box. For datasets where original annotations contained multiple species (Anuraset, Carrion Crows, Hawaiian Birds, Katydid, Powdermill), we chose one species per file to be the positive sound event. For datasets where original annotations contained multiple call types (Marmoset, Rana Sierrae, Jumping Spider), we chose one call type per file to be the positive sound event. The species/call types that were not chosen for that file were considered as background in that file and discarded. Events marked as ``Unknown'' were evaluated as in~\cite{NOLASCO2023Learning}. Unknown events were cropped out of \method prompts.

\subsubsection{Anuraset} Subset of twelve recordings from the strongly-labeled portion of AnuraSet~\cite{canas2023dataset}. The original study collected soundscape recordings from omni-directional microphones placed near four bodies of water in Brazil. The dataset was developed to improve automated classification of anuran vocalizations. Expert annotators identified the start- and end-times of vocalizations, for each of 42 different frog species. For our purpose, we used one frog species per file as positive event (\textit{Boana lundii}, \textit{Leptodactylus latrans}, \textit{Physalaemus albonotatus}, four files each); annotations in these files that corresponded to other species were discarded.

\subsubsection{Carrion Crow} Set of ten hour-long recordings of carrion crows (\textit{Corvus corone}) near Le\'on, Spain. Recordings were made using on-body recorders
attached to the tails of adult crows. Recordings were made through a study investigating communication and cooperative behavior in groups of crows. One expert annotator 
identified the start- and end-times of vocalizations of adult crows and brood-parasitic great spotted cuckoo (\textit{Clamator glandarius}) chicks. We used one species per file as the positive event (five recordings each). Annotations in these files that corresponded to the other species were discarded. Vocalizations by crow chicks, such as begging calls, are considered as background sound. Crow vocalizations were marked as ``Unknown'' when it was difficult to discern the life stage of the vocalizing individual.

\subsubsection{Gunshot} Set of seven recordings taken from the Test split of the gunshot detection dataset presented in~\cite{YOH2024gunshots}. Recordings were made in forests in Gabon, using omni-directional microphones. The dataset was developed to investigate impacts of hunting on biodiversity in Gabon. Annotators marked the start- and end-times of gunshots. For our purpose, we collated all recordings from a single site into a single file. Then, we discarded files which had fewer than seven detected gunshots. In each file, positive events are gunshot sounds.

\subsubsection{Hawaiian Birds} Subset of twelve recordings from the dataset presented in~\cite{navine2022soundscape}. Recordings were made at a variety of locations in Hawaii, using omni-directional microphones. The recordings were collected for a variety of studies conducted by the Listening Observatory for Hawaiian Ecosystems at the University of Hawai‘i at Hilo. Expert annotators were asked to draw a spectrogram bounding box around each vocalization of 27 bird species present in Hawaii. Vocalizations separated by less than 0.5 seconds were allowed to be included in a single bounding box. For our purpose, we used one species per file as the positive event (\textit{Chlorodrepanis virens}, \textit{Myadestes obscurus}, \textit{Pterodroma sandwichensis}, four files each); annotations in these files that corresponded to other species were discarded.

\subsubsection{Hainan Gibbon} Set of nine recordings from the Test split in \cite{dufourq2021automated}. Soundscape recordings were made in Hainan, China, using omni-directional microphones. The dataset was developed to improve monitoring of Hainan Gibbons (\textit{Nomascus hainanus}), a critically endangered primate species. Expert annotators identified gibbon vocalizations, and annotated start- and end-times up to the closest second. For all nine files, positive events are gibbon vocalizations

\subsubsection{Humpback Whale} Subset of ten hour-long recordings from the dataset presented in~\cite{Allen2021}. Recordings were collected at sites in the North Pacific by bottom-mounted, omni-directional hydrophones. The dataset was developed to train a humpback whale (\textit{Megaptera novaeangliae}) vocalization detector. We considered the ``initial'' audit portion of the data from this publication, in which experts annotated full recordings for humpback whale vocalizations and un-annotated time periods implicitly do not contain whale vocalizations. 
For our purpose, we selected ten recordings. The recordings were divided into 75-second chunks, and for each recording we discarded all subchunks which did not contain at least one whale vocalization. In each of the ten files that resulted, the positive events are humpback whale sounds.

\subsubsection{Jumping Spider} Four recordings provided by the Damian Elias lab. Male jumping spiders (\textit{Habronattus} species) perform solid-borne acoustic displays in mating contexts. These displays were recorded using a laser vibrometer directed at the substrate on which spiders were standing. The dataset was collected as part of a study investigating signal evolution across the genus. Expert annotators labeled the start- and end-times of each spider signal, along with a signal-type category from a pre-defined list. For our purposes, we used one signal type per file as the positive event (``thumping'', ``knocking'', two files each). Other signal type annotations were discarded

\subsubsection{Katydid} Subset of twelve recordings from the dataset presented in~\cite{symes2023neotropical}. Recordings were made in the forest canopy of Barro Colorado Island in Panam\'a, using an omni-directional microphone. The dataset was developed to quantify the calling activity of katydids (\textit{Tettigoniidae}). Expert annotators identified the start- and end-times of katydid calls, for 24 species of katydids. For our purpose, we used one species per file as the positive event (\textit{Anaulacomera darwinii}, \textit{Thamnobates subfalcata}, \textit{Pristonotus tuberosus}, four files each); annotations in these files that corresponded to other species were discarded. Additionally, following the original study, we only retained annotations where the annotators were able to identify a clear pulse structure.

\subsubsection{Marmoset} Subset of ten recordings from the dataset presented in~\cite{zhang2018automatic}. Ten juvenile common marmosets (\textit{Callithrix jacchus}) were each placed in a sound-proofed room, away from other individuals, and their spontaneous vocalizations were recorded using a cardioid microphone. The dataset was developed in order to investigate the use of deep learning for detecting and classifying marmoset vocalizations. Annotators identified the start- and end-time of each vocalization, and categorized each vocalization according to one of ten pre-defined call types. For our purpose, we selected one ten-minute file from each individual. In five of these files, positive events are ``Phee'' calls; in the other five files positive events are ``Twitter'' calls. Similar call types (``Peep'', ``Trillphee'', ``Pheecry'', ``TrillTwitter'', ``PheeTwitter'') were re-labeled as ``Unknown'', and the remaining annotated call types were discarded.

\subsubsection{Powdermill} All recordings from the dataset presented in~\cite{Chronister2021Powdermill}. Four dawn chorus soundscapes were captured using omni-directional microphones at the Powdermill Nature Reserve, Pennsylvania, USA. The dataset was developed in order to provide a resource for research into automated bird sound classification and detection in soundscape recordings. Expert annotators marked the start- and end-times of vocalizations from 48 bird species. For our purpose, we used one species per file as the positive event (\textit{Pipilo erythrophthalmus}, \textit{Geothlypis trichas}, two recordings each). Annotations corresponding to other species were discarded.

\subsubsection{Ruffed Grouse} Recordings from the dataset presented in~\cite{Lapp2023rg}. Recordings were made using omni-directional microphones placed in regenerating timber harvests in Pennsylvania, USA. The dataset was developed to evaluate the performance of an automated method to detect ruffed grouse (\textit{Bonasa umbellus}) drumming events. In the original study, five-minute clips were extracted from the original recordings, and annotators marked the start- and end-times of each drumming event. For our purpose, for each of the two months in the recording period (April and May, 2020), we concatenated all the recordings into a single audio file. In each file, the positive sound event is ruffed grouse drumming.

\subsubsection{Rana Sierrae} Subset of recordings from the dataset presented in~\cite{Lapp2024rana}. Underwater soundscapes were captured using omni-directional microphones placed in waterproof cases that were attached to the bottom of a lake in California's Sierra Nevada mountains. The data were collected to characterize the vocal activity of a wild population of the endangered Sierra Nevada yellow-legged frog (\textit{Rana sierrae}). For each vocalization, annotators marked its start- and stop-time, and classified it into one of five call types. For our purposes, we concatenated into one file all recordings from each single day presented in the original dataset. This yielded seven files, corresponding to the seven days of recording. For four files, the positive sound event was one call type, ``Primary vocalization''. For the other three files, the positive sound event was a different call type, ``Frequency-modulated call''. Other call type annotations were discarded.

\subsubsection{Right Whale} Subset of ten recordings from the dataset B$^*$ presented in~\cite{kirsebom2020performance}. Underwater soundscapes were recorded by hydrophones 
moored 5-50 meters above the bottom of the seafloor. The data were originally recorded as part of a study (Simard et al. 2019) documenting changes in the distribution of the endangered North Atlantic right whale (\textit{Eubalaena glacialis}, NARW). Expert annotators manually midpoints of NARW upcalls. For our purpose, we extended each midpoint to a 1-second bounding box. The duration of this box was chosen based on the description of the NARW upcall in~\cite{kirsebom2020performance} as a ``1-s, 100–200 Hz chirp with a 610 Hz bandwidth.'' In each file, the positive sound event is NARW upcall.

\subsection{Data Generation Parameters}

\subsubsection{Sampling}

For sampling target pseudovox, an event rate $r$ is drawn from $\{1, 0.5, 0.25, 0.125,0.0625\}$ events per second. The number $n$ of pseudovox that will appear in a $\operatorname{dur}$-second scene is drawn from a Poisson distribution with rate parameter $r\times \operatorname{dur}$. To reduce the number of event-less scenes, we set $n=\max(n,1)$ with probability $1$ for support scenes and with probability $0.5$ for query scenes. A clustering level $k$ is drawn from $k\in K = \{\left \lfloor{M/128}\right \rfloor ,\left \lfloor{M/64}\right \rfloor,\left \lfloor{M/32}\right \rfloor,\left \lfloor{M/16}\right \rfloor,\left \lfloor{M/8}\right \rfloor\},$ where $M$ is the total number of pseudovox. A cluster $c_T$ is drawn at random from the $k$-means clustering of the pseudovox at this level, and $n$ pseudovox are drawn from this cluster $c_T$. This process is repeated for the distractor cluster $c_D$ and distractor pseudovox.

For background audio, two background tracks are sampled, looped to the scene duration, and overlaid. When constructing support-query pairs, for the query scene these background tracks are different from the support scene background tracks with probability $p_{gen} = 0.5$

\subsubsection{Augmentation}

With probability $0.2$, all target pseudovox are flipped in time. To apply pitch/time shifting, a resampling rate $\rho$ is drawn from $\{0.3, 0.5, 0.7, 1, 1, 1, 1.5, 2\}$ and all target pseudovox are resampled from $16kHz$ to $\rho\times 16kHz.$ For amplitude augmentation, to simulate one or more individuals making sounds at different amplitudes, we construct a random Gaussian mixture model (GMM) with two components. Each component has a mean amplitude $\mu_a\sim \operatorname{Unif}(-12,7)$ dB and a standard deviation $\sigma_a\sim\operatorname{Unif}(0,5)$ dB. The weight of the second mixture component is drawn from $\{0,0,0,0,0,0.1,0.2,0.3,0.4,0.5\}$. The SNR of each target pseudovox is set based on draws from this GMM and set based on the RMS amplitude of the background audio and pseudovox. To add reverb, we convolve with a recorded room impulse response~\cite{rir_data}.
This process is repeated for distractor pseudovox. We apply only resampling augmentations to background audio.

\subsubsection{Combination}

To simulate one or more individuals making sounds at different rates, we construct a random GMM with two components that are used to sample time gaps between events in a scene. Each component has a mean $\mu_t\sim \operatorname{Unif}(0,30)$ seconds and a standard deviation $\sigma_t\sim\operatorname{Unif}(0,10)$ seconds. The weight of the second mixture component is drawn from $\{0,0,0,0,0,0.1,0.2,0.3,0.4,0.5\}$. Target pseudovox are added to the background audio, one-by-one, with timegaps between consecutive events sampled from this GMM. Events that extend past the duration of the scene are looped back to the beginning.
This process is repeated for distractor pseudovox.

\subsubsection{Variations for additional experiments}

 High homogeneity in events: Set cluster level $k=\lfloor{M/8}\rfloor$.
 Low homogeneity in events: Set cluster level $k=\lfloor{M/128}\rfloor$.
 High events / second: Set rate $r=1$ event per second.
 Low events / second: Set rate $r=0.0625$ events per second.
 Only high SNR events: SNR mean $\mu_a$ is drawn from $\operatorname{Unif}(2,7)$ dB.
 Only high SNR events: SNR mean $\mu_a$ is drawn from $\operatorname{Unif}(-12,-7)$ dB.

\newpage

\subsection{Results on validation datasets}

\begin{table*}[h!]
\caption{\textbf{F1} scores @0.3 IoU on \textbf{validation datasets}. Models marked with $\dagger$ were pre-trained using our generated data, rather than the data used in the original publication.}
\label{sample-table}
\begin{center}
\begin{small}
\begin{tabular}{l||cccccc|c}
\toprule
\multicolumn{8}{c}{\textbf{5-shot, within-recording}} \\
\midrule

Model & HB & ME & PB & PB24 & PW & RD & Avg \\
\midrule
BEATs + Linear & .839 & .310 & .067 & .104 & .668 & .159 &.358\\

AVES + Linear  &\textbf{.872} &.305 &.566 &.688 &.578 &.380 &.565  \\

Protonet & .788 & .597 & .321 & .492 & .211 & .359&.461 \\

Protonet$^\dagger$ & .775 &.518 &.480 &.482 &.165 &.335&.459  \\

Transductive &.500 &.173 &.210 &.342 &.085 &.143 &.242  \\

SCL & .719 & .691 & .538 & .688 & .080 & .368&.514 \\

SCL+finetuning  &.779 &.634 &.577 &.713 &.077 &.371 &.525  \\

SCL$^\dagger$ & .578 &.429 &.374 &.674 &.090 &.498 &.440  \\

SCL$^\dagger$+finetuning &.749 &.486 &.379 &.659 &.094 &.353 &.453  \\

\method$^\dagger$ (ours) & .659 & \textbf{.829} & \textbf{.657} & \textbf{.809} & \textbf{.738} & \textbf{.532} &\textbf{.704}\\

\bottomrule
\end{tabular}
\end{small}
\end{center}
\end{table*}

\subsection{\benchmark visualizations}

\begin{figure*}[h!]
\begin{center}
\includegraphics[width=\linewidth]{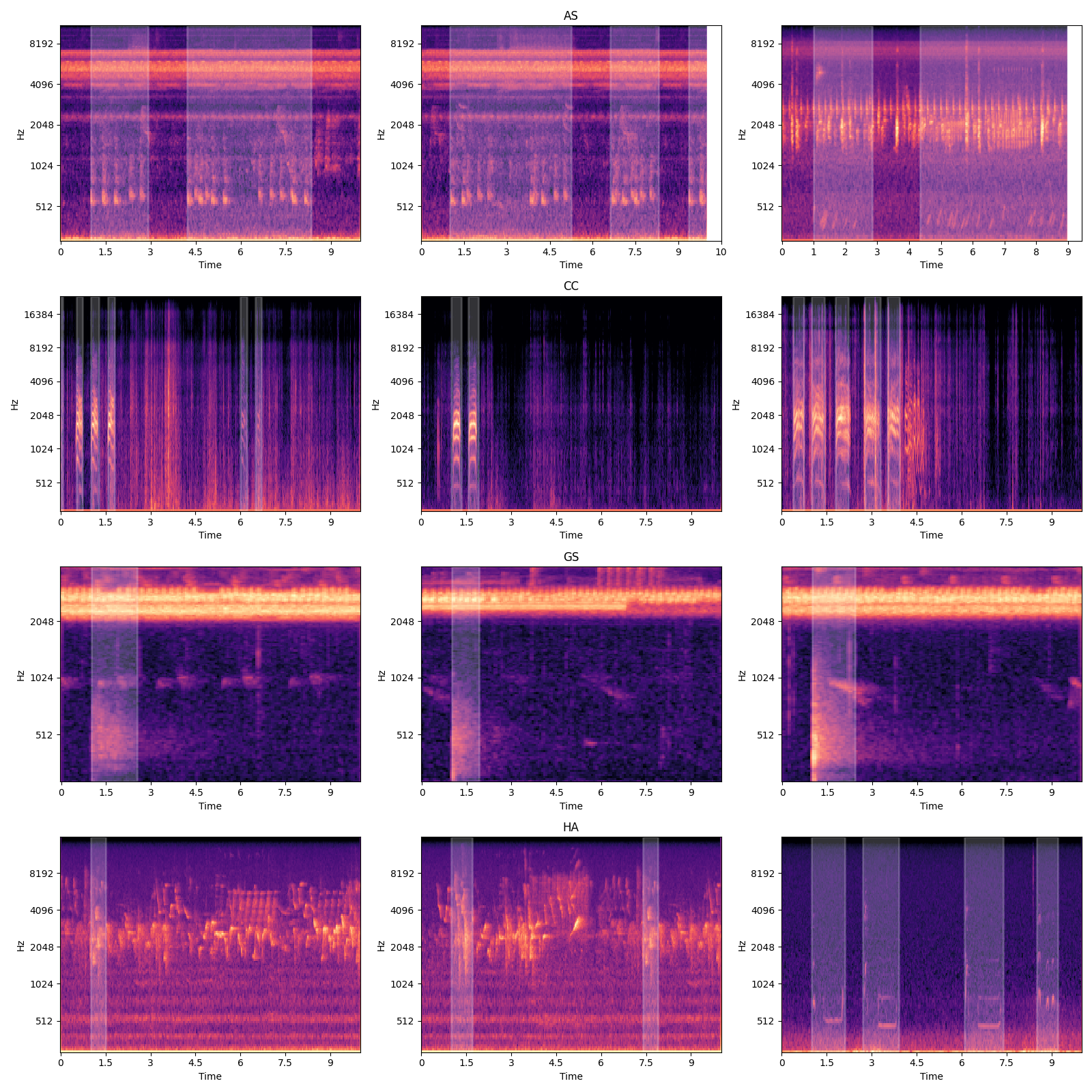}
\caption{Example spectrograms from \benchmark, part 1. Each row contains three spectrograms from one dataset. Positive events are highlighted. AS: AnuraSet, CC: Carrion Crow, GS: Gunshot, HA: Hawaiian Birds.}
\label{spec0}
\end{center}
\end{figure*}

\newpage

\begin{figure*}[h!]
\begin{center}
\includegraphics[width=\linewidth]{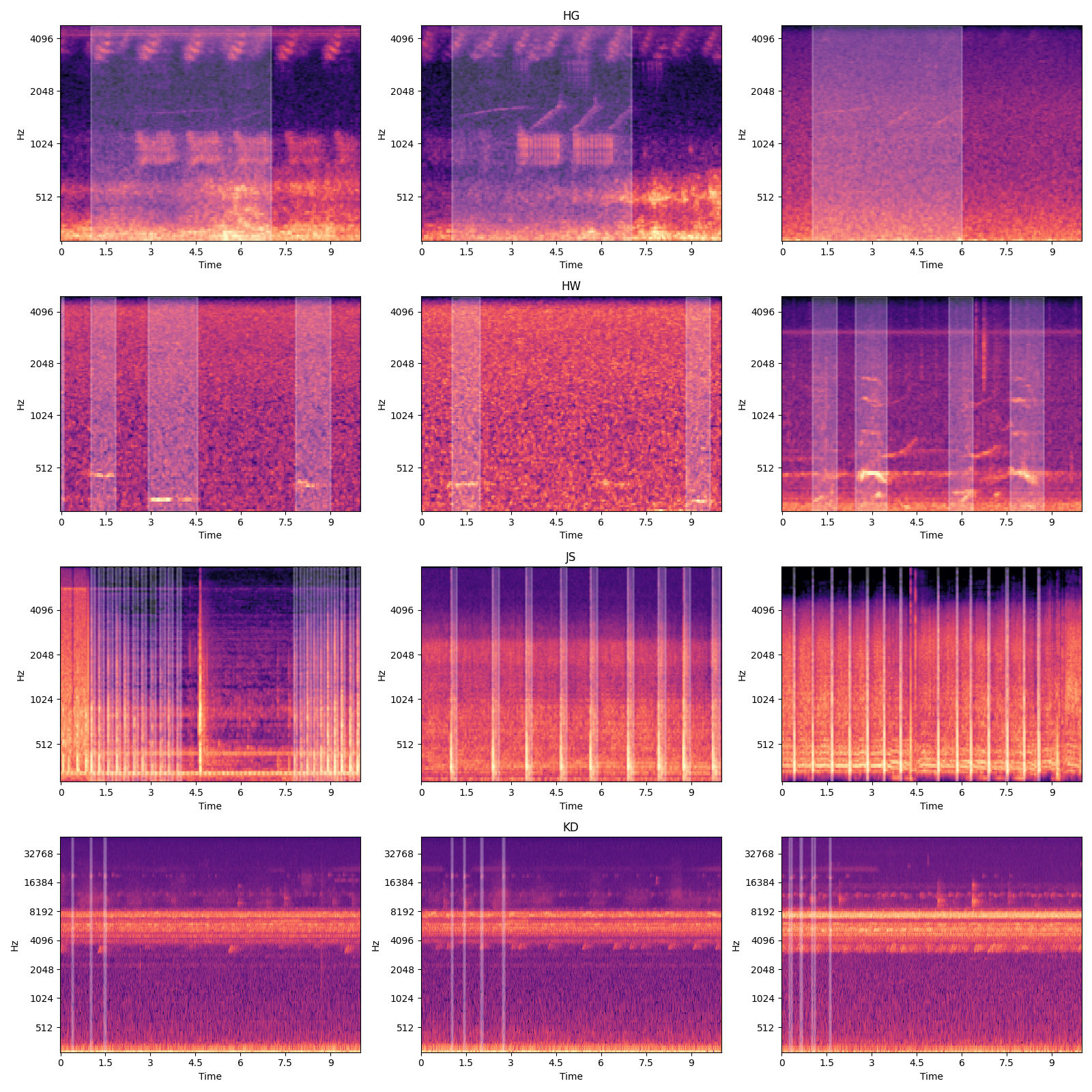}
\caption{Example spectrograms from \benchmark, part 2. Each row contains three spectrograms from one dataset. Positive events are highlighted. HG: Hainan Gibbons, HW: Humpback Whale, JS: Jumping Spider, KD: Katydid.}
\label{spec1}
\end{center}
\end{figure*}

\newpage

\begin{figure*}[h!]
\begin{center}
\includegraphics[width=\linewidth]{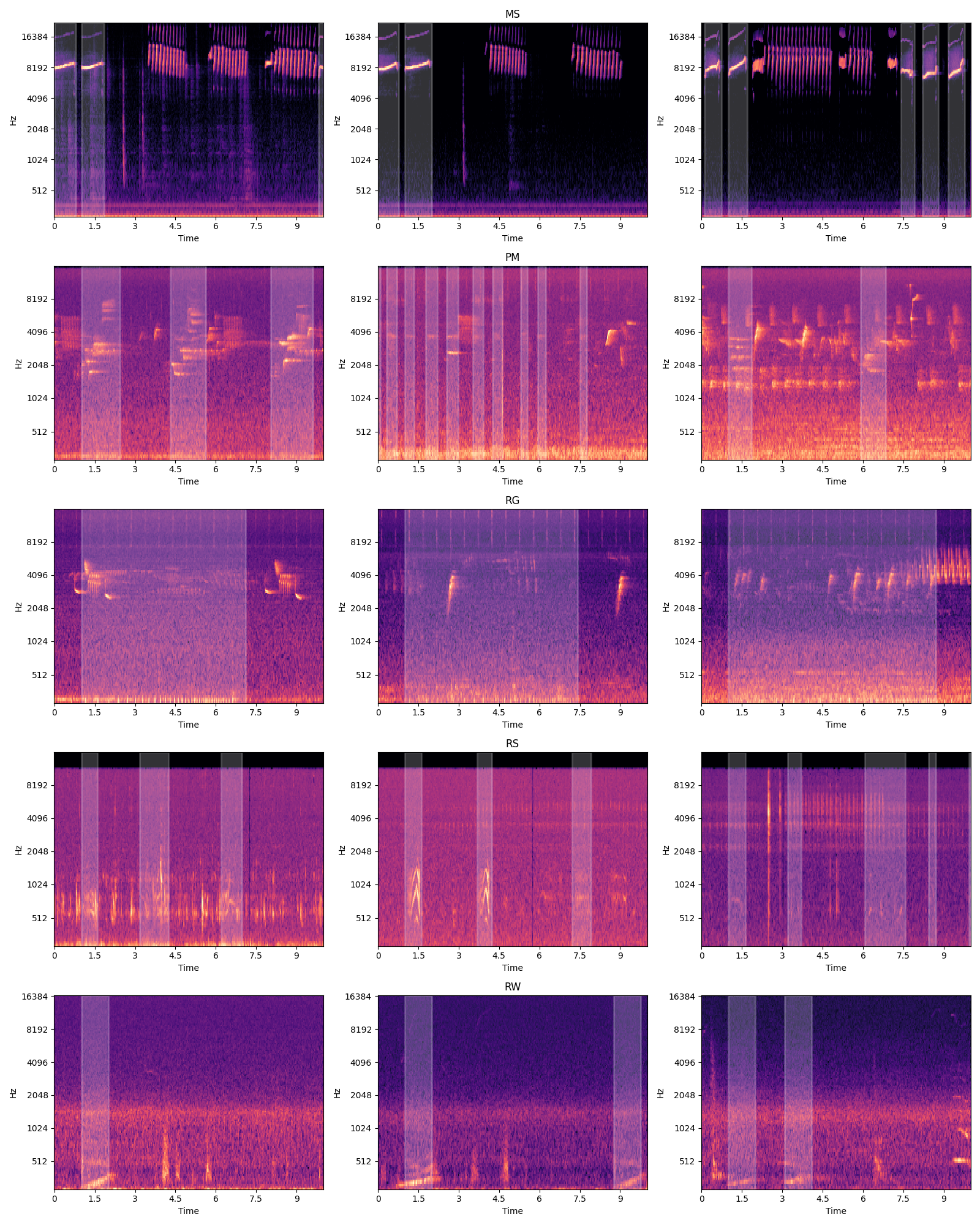}
\caption{Example spectrograms from \benchmark, part 3. Each row contains three spectrograms from one dataset. Positive events are highlighted. MS: Marmoset, PM: Powdermill, RG: Ruffed Grouse, RS: Rana Sierrae, RW: Right Whale.}
\label{spec2}
\end{center}
\end{figure*}


\end{document}